# Symmorphic intersecting nodal rings in semiconducting layers

Cheng Gong[1], Yuee Xie[1], Yuanping Chen[1]*, Heung-sik Kim[2] and David Vanderbilt[2]

[1]*School of Physics and Optoelectronics, Xiangtan University, Xiangtan, 411105, Hunan, China*
[2]*Department of Physics and Astronomy, Rutgers University, Piscataway, New Jersey 08854-8019, USA*

The unique properties of topological semimetals have strongly driven efforts to seek for new topological phases and related materials. Here, we identify a critical condition for the existence of intersecting nodal rings (INRs) in symmorphic crystals, and further classify all possible kinds of INRs which can be obtained in the layered semiconductors with Amm2 and Cmmm space group symmetries. Several honeycomb structures are suggested to be topological INR semimetals, including layered and "hidden" layered structures. Transitions between the three types of INRs, named as α-, β- and γ-type, can be driven by external strains in these structures. The resulting surface states and Landau-level structures, more complicated than those resulting from a simple nodal loop, are also discussed.

Corresponding author: chenyp@xtu.edu.cn

Following the experimental detection of Fermi-arc surface states in Weyl semimetals[1-3], considerable attention has focused on the investigation of topological semimetals/metals (TMs) more generally[4-8]. Classic examples of TMs are the Weyl and Dirac semimetals[9-15], which exhibit twofold and fourfold degenerate Fermi points respectively. These nodal-point semimetals display a number of exotic transport phenomena such as negative magnetoresistance and the chiral magnetic effect[16-19]. Nodal line/ring semimetals belong to another class of TMs[20-27], in which the valence and conduction bands cross along one-dimensional lines in three-dimensional (3D) k-space. In general, the line is not pinned at the Fermi energy[28-30], but passes through the Fermi energy at discrete points. As a consequence, the Fermi surface takes the shape of a thin tube with changing radius, possibly with constrictions. These semimetals are expected to exhibit graphene-like Landau levels and enhanced sensitivity to long-range Coulomb interaction[31-36]. Unlike nodal points, nodal lines/rings can form various topologically connected structures such as chains[37,38], knots[39], and Hopf links[40-42], bringing new physics and topological properties.

On the other hand, two-dimensional materials are the focus of another recently thriving field[43]. After graphene, many graphene-like honeycomb structures have been proposed, and some of them have been fabricated successfully[44-49], including silicene, germanene, BN, and phosphorene. These not only show intrinsic interesting properties in single-sheet form, but also have interesting hybrid properties when stacked into 3D materials[50-52]. These stacked structures often have a mirror symmetry along the stacking direction, and since nodal lines/rings can be protected by mirror symmetry[53-56], it is natural to ask whether we can obtain topological nodal line/ring semimetals by the stacking of layered structures.

Here we identify a necessary condition for the existence of intersecting nodal lines in symmorphic crystal structures. By stacking semiconducting honeycomb layers, three types of intersecting nodal rings (INRs), as shown in Fig. 1, are found to occur. The α-type consists of isolated crossed rings as in Panel (a), the β-type corresponds to a nodal chain like that in Panel (c), and the γ-type is the structure of ladder of parallel rings as in Panel (d). Moreover, the three topological phases can be converted into one another via application of external strain. Interesting surface states and Landau levels (LLs) in these INR semimetals are discussed. Several 3D layered or "hidden" layered materials are suggested to possess the topological nodal rings. A tight-binding (TB) model is used to explain the relations between the topological phases and how they evolve into one another.

For the INRs to be protected, a critical necessary condition is *the presence of at least two intersecting mirror or glide planes commuting with each other* in the crystal structure[57]. For simplicity, here we consider only two bands (occupied/unoccupied) near the Fermi level in the presence of two mirror planes without spin-orbit coupling. Let us denote the two mirror planes in the momentum space as *A* and *B*, as shown in Fig. 1(a). The occupied and unoccupied bands on the *AB*-intersecting line, say between points X and Y, can be labeled with two mirror eigenvalues $a^{\pm}$ and $b^{\pm}$ taking values ±1. The right half of Fig. 1(b) shows the bands on the XY-line. If the two bands have eigenvalue pairs $(a^+,b^+)$ and $(a^-,b^-)$ respectively, then they can cross without a gap opening[58]. If we deviate from the XY-line to look at the bands on a generic k-path residing in plane *A* (see the curved arrow in Fig. 1(a)), then the two bands can cross again because of different *A* eigenvalues for each band, as depicted in the left half of Fig. 1(b). The same argument applies to plane *B*, hence this guarantees the presence of two nodal lines in planes *A* and *B* respectively, meeting at the band crossing point on the XY line and forming a nodal link. Note that non-symmorphic (glide or screws) characters are not mandatory, so this kind of nodal chains can exist even in symmorphic crystals[59-61], in contrast to a previous suggestion where the non-symmorphic nature was essential[37]. Whether the nodal lines are closed or open depends on details of the band dispersion, and the α-type INR can be transformed into β- or γ-type as shown from our following results.

According to the necessary condition of INRs, two structure types are considered (Fig. 2). The first kind consists of 3D layered structures with $sp^2$-hybridization atoms, as shown in Fig. 1(a), with the planar layers stacked in an AA' stacking sequence (Fig. 2(c)). Each layer consists of hexagonal rings with each ring including two types of atoms labeled A1 and A2. The four-atom primitive cell (two A1 and two A2) is shown in Fig. 2(b). Atoms of the same type form dimers along the armchair direction, while those of opposite type make up the zigzag chains. The second structure type is a porous network in which $sp^2$-hybridized zigzag chains are connected by $sp^3$-hybridized linker atoms (Fig. 2(d)). Its primitive cell in Fig. 2(e) includes six (two $sp^3$ and four $sp^2$) atoms. Since the bands closest to the Fermi level will be dominated by the $sp^2$ atoms, it is reasonable to neglect the $sp^3$-hybridized atoms in a first approximation, in which case the sp² atoms form a structure of buckled layers stacked in an AA' sequence (Fig. 2(f)). The angle θ between lattice vectors is defined in the figures. Both types of structures have two mirrors on the planes *xz* and *xy*.

When only one orbital of each atom in Fig. 2(a) is considered, a 4×4 TB model can be used to describe its electronic properties:

$$H = \sum_i \varepsilon_\alpha a_i^\dagger a_i + \sum_{i,j} t_\beta a_i^\dagger a_j, \tag{1}$$

where $a_i^\dagger/a_j$ represent the creation/annihilation operators, $\varepsilon_\alpha$ ($\alpha = 1,2$) represent site energies of atoms A1 and A2, $t_\beta$ ($\beta = 1 \ldots 7$) are the hopping parameters between atoms. Here $t_1$ to $t_5$ describe the intra-layer interactions, while $t_6$ and $t_7$ describe the inter-layer couplings (Fig. 2(b)). When the sp³-hybridization atoms in Fig. 2(d) are neglected, the porous network becomes a layered structure. From this point of view, the main difference between the structures of Figs. 2(a) and 2(d) is that the layers in the latter are buckled rather than planar. Because of this close analogy, Eq. (1) can be used to describe the electronic properties of both structures.

We start from a semiconducting single layer. In this case the interlayer interactions in Eq. (1) can be omitted, i.e., we can set $t_6 = t_7 = 0$. The dashed red lines in Fig. 3(a) show the band structure of a typical single-layer semiconductor. It has a substantial band gap, and completely flat bands along paths Z-T, R-T and T-S because of the absence of interlayer couplings. When the semiconducting layers are stacked into a 3D structure, the interlayer couplings $t_6$ and $t_7$ become involved. As a result, the flat bands become dispersive, and the conduction and valence bands cross at the Fermi level. In Fig. 3(a) these crossings look like Dirac points, but as we shall see, they link together in 3D to form nodal rings or lines.

By tuning the parameters in Eq. (1), the three types of INRs in Fig. 1 can be generated. Figure 3(a) presents the band structure for the α-type rings. One can find that there are crossings along Z-T, T-Y, R-T and T-C. In the full Brillouin-zone (BZ), these crossing points lie on two perpendicular nodal rings with a common center at T, as shown in Fig. 3(d). One ring lies on the $k_a = k_b$ plane (plane $A$) while the other lies on the $k_c = 0$ plane (plane $B$). By comparing the band eigenvalues, it can be seen that this pattern corresponds to the α-type phase in Fig. 1(a).

By increasing the intralayer hoppings while decreasing the interlayer ones, the band structure in Fig. 3(a) evolves into that of Fig. 3(b) by inverting occupied/unoccupied bands at C and R points, after which we find crossings along Z-T, T-Y and C-Z. These crossing points lie on two perpendicular INRs on planes $k_a = k_b$ and $k_c = 0$ centered on the points T and Z respectively, as shown in Fig. 3(e). They link in the full BZ and form a nodal chain, corresponding to the β-type phase in Fig. 1(c). This phase is different from the type of nodal chain described in Ref. [36], which is protected by a nonsymmorphic glide-plane symmetry.

By contrast, when the intralayer hoppings are decreased while the interlayer ones are increased, the band structure in Fig. 3(a) evolves into that of Fig. 3(c) by inducing a band inversion at Z. This introduces an additional nodal ring on the $A$ plane encircling Z, and the ring on the $B$ plane is now open and connects the two rings on plane $A$ as shown in Fig. 3(f). The crossing points are now located on the k paths Γ-Z, T-Y, R-T, T-C and C-Z. In the extended BZ the nodal structure has an appearance like a ladder of parallel rings, corresponding to the γ-type phase in Fig. 1(d). The topological protection of the three types of INRs can also be inferred from their 1D winding numbers along a close path $\mathcal{L}$ encircling the rings: $N_{\mathcal{L}} = \frac{1}{\pi}\oint_{\mathcal{L}} d\bm{k} \cdot A(\bm{k})$, where $A(\bm{k})$ is the Berry connection at the point $\bm{k}$. The calculation results indicate that all of them have nontrivial values.

To find topological materials possessing these INR phases, we construct structures like Figs. 2(a) and 2(d) based on IV or III/V elements. By calculating band structures using density functional theory (DFT) [62], we find that layered structures BN, AlP and GaP and "hidden" layered structures SiC, BP and BAs can fit the requirements (Fig. S7 in SI). The structural parameters of these structures are shown in Table S1. We calculate their phonon dispersions, and find that there are no soft modes in the spectra of BN and SiC (Fig. S8 in SI). This indicates that BN and SiC are metastable structures having good stability. Therefore, BN and SiC are used as two examples to exhibit the topological properties.

Figure 4(a) shows the band structure of single-layer honeycomb BN, which we find to be a semiconductor with a direct band gap. After the BN layers are stacked into 3D structure by AA' stacking, the band structure changes as shown in Fig. 4(b), which looks quite similar to Fig. 3(a). A close examination indicates that there are indeed α-type nodal rings in BZ. The projections of the band structures illustrate that the states around the Fermi level are contributed mainly by $p_z$ orbitals on B and N atoms. Therefore, it is reasonable that we use Eq. (1) to describe the structures [Detail parameters for fitting the DFT results can be seen in Table S2 in SI].

The band structure of "hidden" layered SiC structure is shown in Fig. 4(c). It is also very similar to the spectrum in Fig. 3(a), and α-type nodal ring is also found here. As mentioned above, the α-type phase evolves into β- and γ-type phases by tuning the hopping energies. An external strain along the direction (110) can induce the same effect as the variation of hopping parameters. As the angle θ changes with the strain from 89º to 80º, meanwhile He atoms are squeezed into the holes of the porous structure[63,64] (the inset in Fig. 4(d)), the band structure changes to that of Fig. 4(d). It is similar to the band spectrum in Fig. 3(c), which means that the system is changed to a β-type nodal-ring semimetal. After θ is increased further from 89º to 108º, the band structure in Fig. 4(e) corresponds to the γ-type nodal ring. As seen from Figs. 4(c-e), all three types of INR structures are accessible for SiC under strain.

Figure 5 presents [010] and [$\bar{1}$10] surface band structures of the three kinds of INRs. On the [010] surface, we find that all types of INRs exhibit drumhead states inside the projections of the nodal rings (Figs. 5(a-c)). However, the surface states on the [$\bar{1}$10] surface are different. The surface states of the α-type phase are still drumhead states, as shown in Fig. 5(d). Instead, in the cases of β- and γ-type nodal rings, the linking of the nodal rings induces exotic surface states. In Fig. 5(e), the surface states are distributed in a dumbbell-like region with the two ends corresponding to overlap regions. In Fig. 5(f), the surface state region has the appearance of a donut or an annular eclipse, because the projection of one ring is right in the center of the other. The areas of the surface state regions can be tuned by strain, and the transition between the three types of linked rings can also be tuned.

In search of other types of possible symmetry-allowed INRs, we apply a band representation analysis for our $p_z$-orbital model [65-73]. Although many different kinds of nodal line structures are allowed, as listed in SI [70], we do find that the $\alpha$- and $\beta$-, and $\gamma$-type INRs we presented here are actually an exhausting set of allowed INRs. This is because the formation of nodal intersecting point is not allowed on another mirror-intersecting line $\Gamma$-Y, due to the absence of irreducible representations necessary to form the INR. We comment that, a similar analysis on various types of nodal line structures in a non-symmorphic crystal was done in Ref. [73].

One interesting consequence of INRs would be the emergence of flat zeroth LLs in the presence of a magnetic field ***B*** applied along the mirror-intersecting line. As discussed in Ref. [74], one should have a set of 2N essentially degenerate zeroth LLs at any given wavevector $k$ along this line, where N is the number of nodal rings spanning the wavevector interval where this $k$ is found. Thus, when there is a chain of β-type INRs connected in k-space, this should yield a flat band of zeroth LLs extending over almost all $k$, with only small LL gaps opening in the vicinity of the nodal intersections. Electron or hole doping, yielding fully filled or empty zeroth LLs, could lead to a rare realization of the 3D quantum Hall effect[74]. In addition, since the density of states of the zeroth LL varies as a function of angle between the B-field and nodal-ring plane, angular magnetoresistance measurements should be useful in distinguishing between different types of nodal rings[75].

In conclusion, we suggest a generic condition for the presence of INRs and classify them in layered semiconductor materials. These INR semimetals show interesting and unique transport properties including topological surface states and LLs. Our results suggest a guiding principle to engineer INR semimetals not only in fermionic systems but also in photonics crystals or other bosonic lattices, shedding light on nodal line engineering for further studies.

We thank Jun Won Rhim for his insightful comments. YPC and YEX were supported by the National Natural Science Foundation of China (Nos. 51376005 and 11474243). DV was supported by NSF DMR-1408838 and HSK was supported by NSF DMREF DMR-1629059.

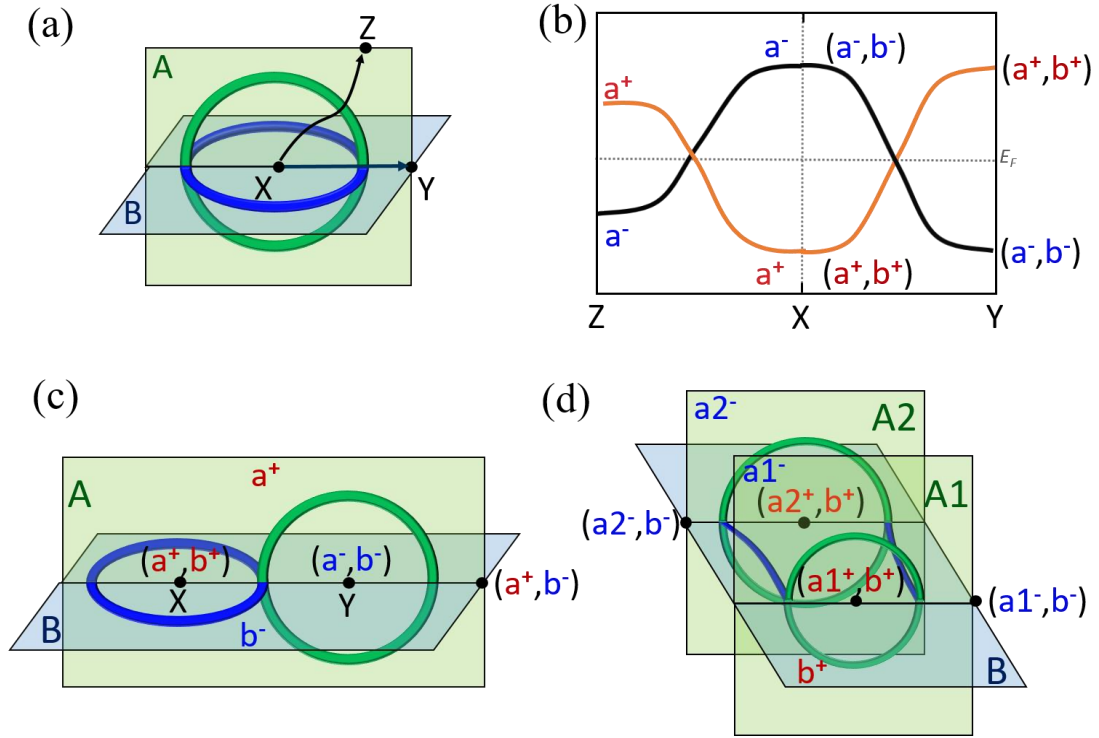

Figure 1. (color online) (a) An α-type INR, and (b) a simple model band structure manifesting nodal links from the two mirror symmetries *A* and *B* in (a). The right and left halves of (b) correspond to the bands on the k-path X-Y (straight arrow in (a)) and X-Z (curved black arrows), respectively. Different color (red and blue) represents different symmetry eigenvalues. (c) β-type and (d) γ-type INRs. Symmetry eigenvalues of the occupied band are shown in (a), (c) and (d).

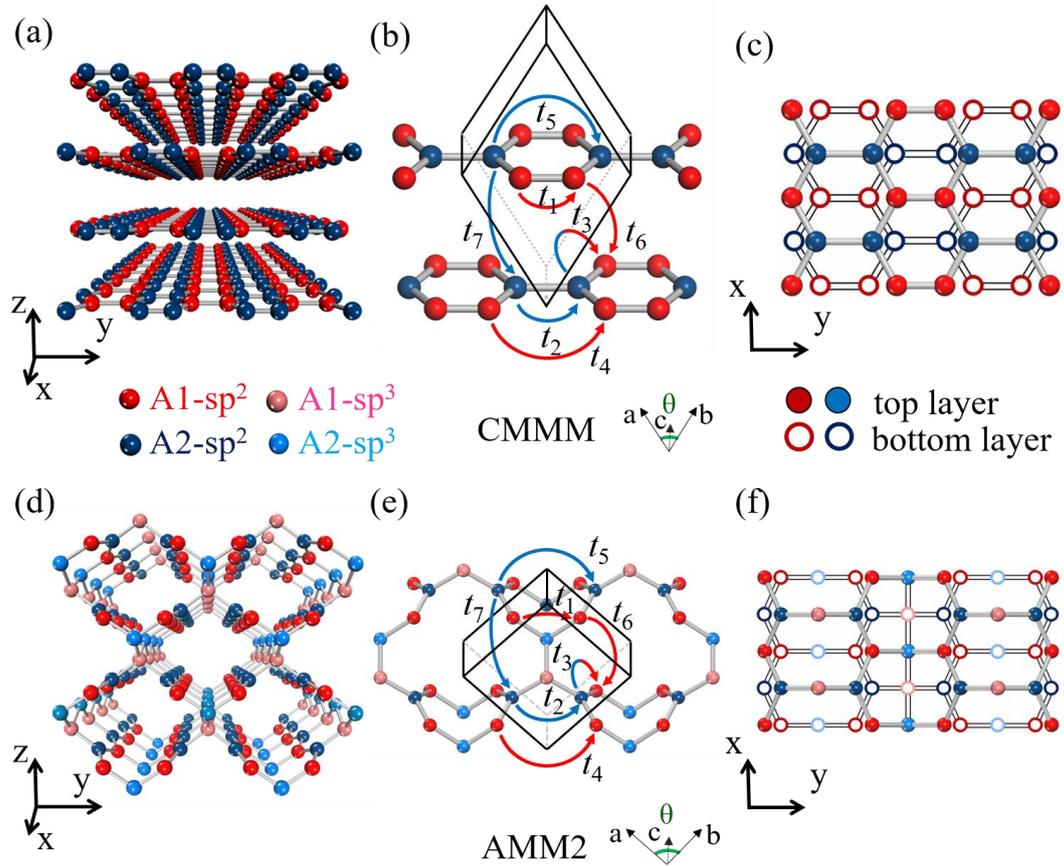

Figure 2. (color online) AA' stacked honeycomb layered structure (a), its primitive cell (b) and top view (c). "Hidden" AA' stacked honeycomb layered network (d), its primitive cell (e) and top view (f). Both of the structures are made of two kinds of atoms A1 and A2. $t_1 \sim t_7$ in (b) and (e) describe the hopping parameters of the structures.

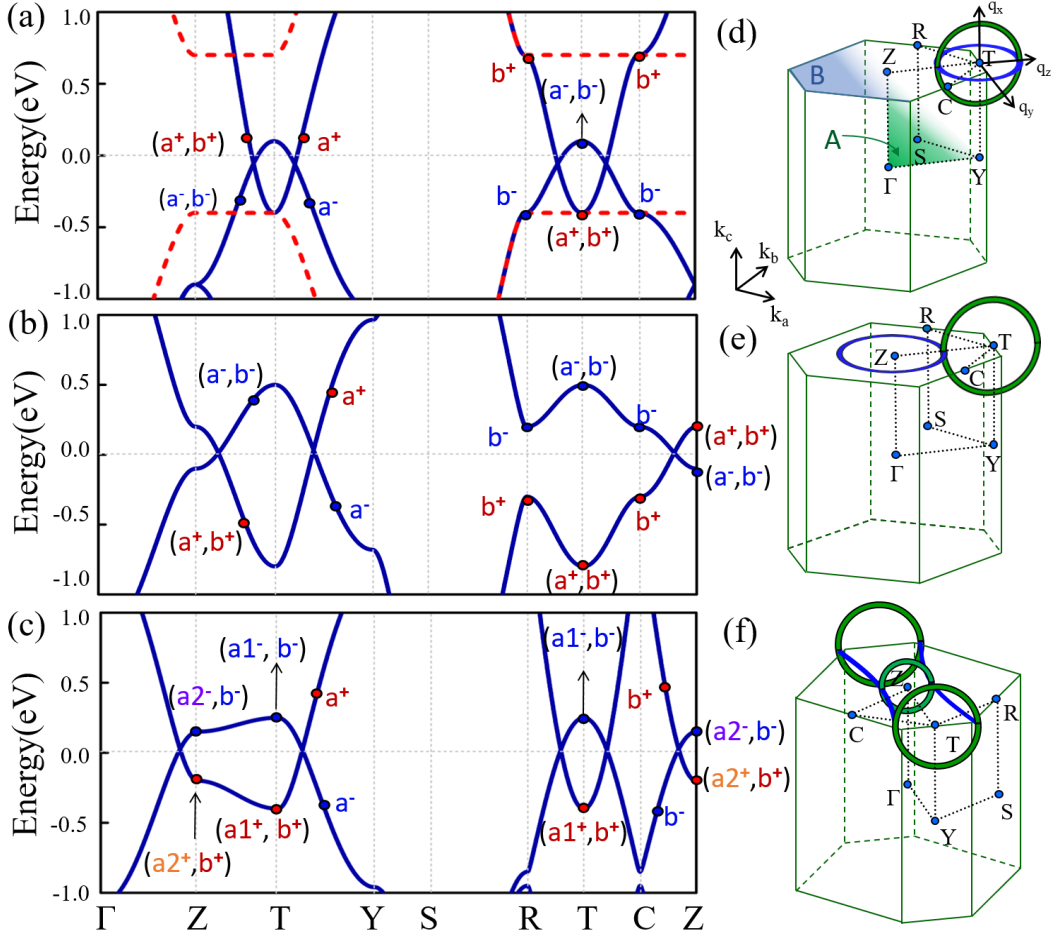

Figure 3. (color online) Band structures based on Eq. (1) with different parameters: (a) $t_1=-1$, $t_2=-0.5$, $t_3=-1.2$, $t_4=t_5=0$, $t_6=0.55$, $t_7=0.25$; (b) $t_1=-2.0$, $t_2=-1.1$, $t_3=-1.2$, $t_4=t_5=0$, $t_6=0.25$, $t_7=0.15$; (c) $t_1=-0.1$, $t_2=-0.05$, $t_3=-1.2$, $t_4=t_5=0$, $t_6=1.0$, $t_7=0.55$. Other parameters are $\varepsilon_1=1.7$, $\varepsilon_2=-0.9$. All the values are in units of eV. Red dashed line in (a) depicts band structure for a single-layer semiconducting with the same parameters as (a) but $t_6=t_7=0$. (d-f) Arrangements of the topological INRs in reciprocal space corresponding to the band structures in (a-c) respectively. In (a-c), eigenvalues for A- and B-mirror planes are shown, where the two mirror planes are illustrated in (d). Note that, in (f), the A-eigenvalues at T and Z are denoted in different symbols (a1$^\pm$ and a2$^\pm$ respectively), and the BZ is different from (d-e) because of changed unit cell parameters, which is used to mimic the DFT results discussed later.

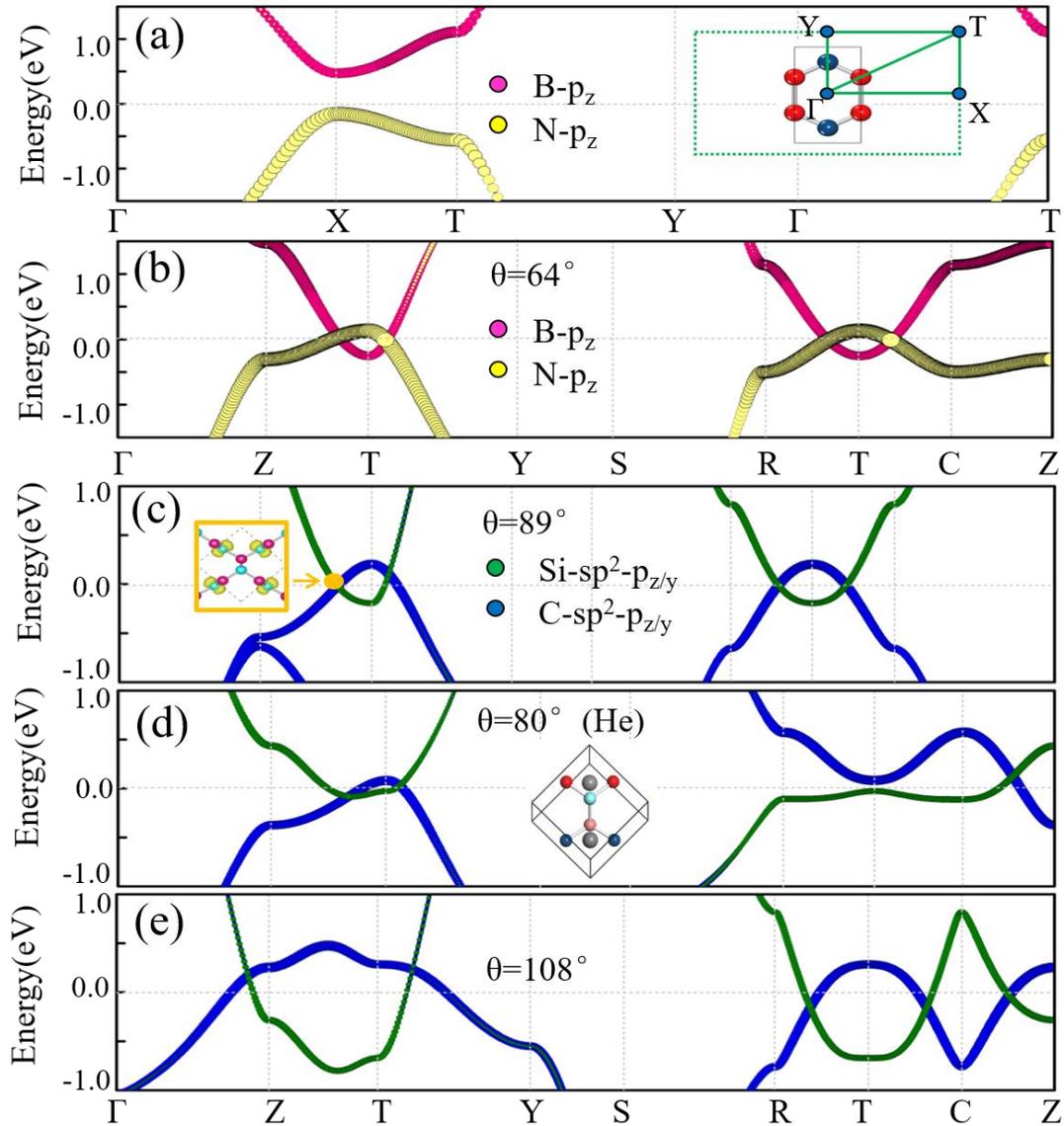

Figure 4. (color online) Projected band structures of (a) single-layer and (b) stacked 3D layered BN (Fig. 1(a)). Projected band structures of "hidden" layered structure SiC with θ = 89⁰ (c), 80⁰ (d) and 108⁰ (e). Insets: (c) charge density of a state around the nodal point, indicating the bonds are similar to the π bonds in graphene; (d) a primitive cell of SiC where He atoms are inserted into the holes.

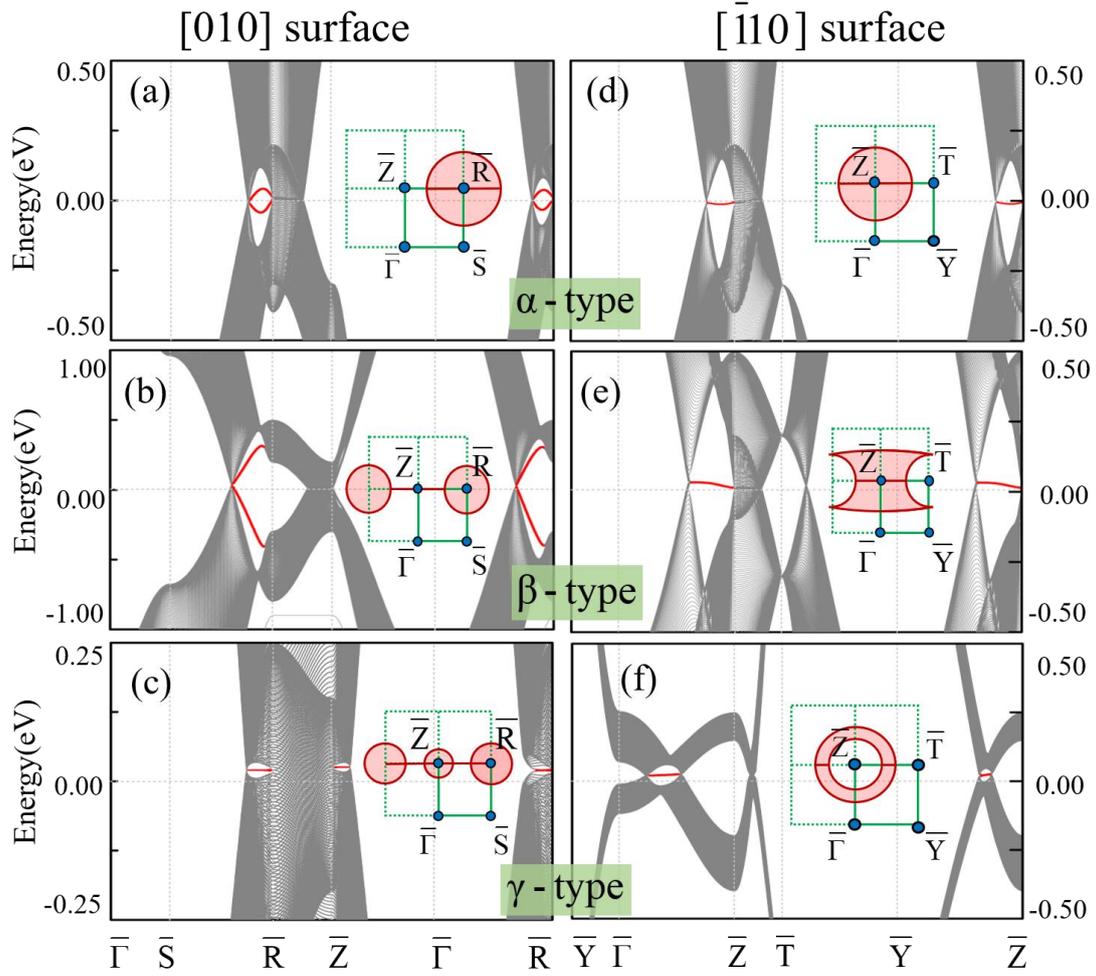

Figure 5. (color online) (a-c) Topological surface states for α-, β- and γ-type LNRs, respectively, on the [$\bar{1}$10] surface. Insets show the surface states regions (red shadows) in the BZ. (d-f) Same but for the [$\bar{1}$10] surface. Because the [010] slabs are terminated by two different surfaces, two different surface states appear in (a-c). However, the surfaces of [$\bar{1}$10] slabs are the same, and thus the two surface states in (d-f) are degenerate.